# Dark matter in elliptical galaxies

C.M. Carollo,[1,2] P.T. de Zeeuw,[1] R.P. van der Marel,[1,3,4] I.J. Danziger,[2] and E.E. Qian[1,5]

[1]Sterrewacht Leiden, Postbus 9513, 2300 RA Leiden, The Netherlands

[2]European Southern Observatory, Karl–Schwarzschildstrasse 2, Garching bei München, Germany

[3]Institute for Advanced Study, Olden Lane, Princeton, NJ 08540

[4]Hubble Fellow

[5]Department of Mathematics, Massachusetts Institute of Technology, Cambridge, MA 02139

# ABSTRACT

We present measurements of the shape of the stellar line–of–sight velocity distribution out to two effective radii along the major axes of the four elliptical galaxies NGC 2434, 2663, 3706, and 5018. The velocity dispersion profiles are flat or decline gently with radius. We compare the data to the predictions of $f = f(E, L_z)$ axisymmetric models with and without dark matter. Strong tangential anisotropy is ruled out at large radii. We conclude from our measurements that massive dark halos must be present in three of the four galaxies, while for the fourth galaxy (NGC 2663) the case is inconclusive.

*Subject headings:* dark matter – galaxies: elliptical and lenticular, cD – galaxies: individual: NGC 2434, NGC 2663, NGC 3706, and NGC 5018 – galaxies: kinematics and dynamics – galaxies: structure

## 1. Introduction

The presence of dark matter in normal elliptical galaxies is still controversial. Although none seems to be required inside the half–light radius $R_e$, an extended halo of dark matter might surround the luminous regions (e.g., de Zeeuw & Franx 1991; Bertin & Stiavelli 1993). Evidence for such dark halos has been found for some galaxies through studies of HI kinematics (e.g., Franx, van Gorkom & de Zeeuw 1994), X–ray emission (e.g., Jones et al. 1994), radial velocities of globular clusters (e.g., Mould et al. 1990) and planetary nebulae (e.g., Hui et al. 1994), or gravitational lensing (e.g., Maoz & Rix 1993).

The shallow decline of the stellar line–of–sight velocity dispersion $\sigma$ seen in many normal ellipticals has also been taken as evidence for unseen mass at large radii (Saglia et al. 1993; Bertin et al. 1994; Carollo & Danziger 1994a,b, hereafter CDa,b). Unfortunately, a stellar orbital structure which becomes biased towards tangential motion at large radii can give rise to an almost flat $\sigma$ profile as well, and hence can mimic the presence of dark matter. This ambiguity in the interpretation of stellar kinematic data can be resolved by measuring the *shape* of the stellar velocity distribution along the line of sight (hereafter 'velocity profile', VP) at large radii, since this depends strongly on the orbital anisotropy (Dejonghe 1987; Gerhard 1991, 1993).

Accurate VP measurements in the outer, faint regions of galaxies are not easy to achieve, because a signal–to–noise ratio S/N $\geq 50$ per Å is required to measure the lowest order anti-symmetric and symmetric deviations of the VP from a pure Gaussian (e.g., Rix & White 1992; van der Marel & Franx 1993, hereafter MF). Accurate VP shape measurements were obtained recently by, e.g., Bender, Saglia & Gerhard (1994) and van der Marel et al. (1994a, hereafter MRCFWZ), but these were restricted to radii $\lesssim \frac{1}{2}R_e$.

In this *Letter* we present VP shape measurements along the photometric major axes of the four galaxies NGC 2434, 2663, 3706, and 5018, classified as E in the RC3 (de Vaucouleurs et al. 1991). They cover a range in luminosity, environment, metallicity and rotational properties. NGC 3706 and 5018 have faint shells (Malin & Carter 1983). The measurements extend to radii $\gtrsim 2R_e$, and allow us to quantify the anisotropy of the stellar velocity distribution and to determine whether dark halos are present.

## 2. Velocity profile measurements

Standard $R$ ESO–Bessel broad–band imaging and deeply–exposed major axis long–slit spectra were taken for NGC 2434 and 3706 at the 2.2m telescope (EFOSC2; 4600–6000Å; 5Å FWHM; exposure times 5 and 8 hrs., respectively) and for NGC 2663 and 5018 at the New Technology Telescope (EMMI; 4700–5500Å; 4.5Å FWHM; exposure times 6 and 5 hrs., respectively), at ESO, La Silla. Detailed information about these data and their reduction is given in CDa,b, who measured the $R$-band surface brightness profiles and derived the mean line–of–sight velocity $V$ and velocity dispersion $\sigma$ as a function of projected radius, based on the assumption of a Gaussian VP.

We reanalyzed the spectra (rebinned spatially to have $S/N > 50$ per Å at all radii) with the Fourier–Fitting package of Franx, Illingworth & Heckman (1989), as modified by MF to allow for non–Gaussian VP shapes. The VP is expanded into an orthogonal Gauss–Hermite series with parameters: (i) the line strength, mean $V$, and dispersion $\sigma$ of the best-fitting Gaussian VP; and (ii) the Gauss-Hermite moments $h_3, h_4$ and higher orders that measure deviations from the best-fitting Gaussian. The parameter $h_3$ measures the lowest order anti-symmetric deviation from a Gaussian, $h_4$ the lowest order symmetric deviation. A value $h_4 > 0$ generally indicates that the VP is more centrally peaked than a Gaussian, a value $h_4 < 0$ that it is more flat-topped. The $S/N$ and instrumental resolution of our data are insufficient to measure higher order coefficients such as $h_5$ and $h_6$.

The Fourier–Fitting program is based on an algorithm that minimizes the $\chi^2$ in the Fourier domain. For each galaxy, one template spectrum was constructed as a weighted mix of individual stellar spectra. The weights were chosen to optimize the fit to the observed galaxy spectra, using a technique similar to that of Rix & White (1992). Tests indicate that systematic errors in the VP parameters due to residual template mismatching and imperfect continuum subtraction are negligible for $h_3$. For $h_4$ they can be of order $\pm 0.03$, but they are generally independent of radius.

The resulting radial profiles of $V$, $\sigma$, $h_3$, and $h_4$ for the four galaxies are shown in Figure 1. NGC 2434 has a rotating core, but shows negligible rotation beyond $5''$. The $\sigma$ profile has a central peak and is flat beyond $5''$. NGC 2663 has no or very small rotation, and has a steeply declining $\sigma$ profile. NGC 3706 and 5018 are fast rotators. NGC 3706 has a central peak in $\sigma$, while NGC 5018 has a central minimum. For both galaxies, the $\sigma$ profile shows a modest decline in the outer parts. The measured velocity dispersions for all galaxies are significantly larger than the instrumental dispersion of $\sim 130\,\mathrm{km\,s^{-1}}$.

The $V$ and $h_3$ profiles generally have similar radial shape but opposite sign, in agreement with other studies (MF; Bender et al. 1994; MRCFWZ). In the outer parts of NGC 3706, where $V/\sigma$ is large, $h_3$ and $V$ have the same sign. The VP asymmetries measured by $h_3$ are caused by the intrinsic skewness of the local stellar velocity distribution and by line–of–sight projection (e.g., Evans & de Zeeuw 1994). The $h_4$ profiles differ considerably among the four galaxies. Out to $\gtrsim 2R_e$, $h_4 \gtrsim 0$ for NGC 2434 and 5018, $h_4 \gtrsim -0.05$ for NGC 3706 and $h_4 \gtrsim -0.1$ for NGC 2663.

## 3.  Dynamical modeling

We compare the data to axisymmetric dynamical models in which the phase-space distribution function (DF) $f$ depends only on the two classical integrals of motion, the energy $E$ and angular momentum $L_z$ along the symmetry axis. The even part $f_e(E, L_z) \equiv$ $[f(E, L_z) + f(E, -L_z)]/2$ is determined uniquely by the luminous mass density $\rho_L(R, z)$, for any potential $\Psi(R, z)$ ($R$, $\phi$ and $z$ being the usual cylindrical coordinates). The odd part $f_o$ can be freely specified, provided that $f \equiv f_e + f_o \geq 0$ for all physical values of $(E, L_z)$. Methods for constructing $f(E, L_z)$ DFs for realistic mass distributions were developed only recently (Hunter & Qian 1993; Dehnen & Gerhard 1994; Qian et al. 1994, hereafter QZMH).

The even and odd parts of the DF determine the even and odd parts of the VPs, respectively: $\mathrm{VP}_{e,o}(v) \equiv [\mathrm{VP}(v) \pm \mathrm{VP}(-v)]/2 \equiv (1/\Sigma) \iiint f_{e,o}(E, L_z) \, dv_{x'} \, dv_{y'} \, dz'$, where $\Sigma$ is the projected surface mass density, $(x', y')$ are on the plane of the sky, and $z'$ is along the line of sight. Since we are interested in the mass distribution, we consider only $\mathrm{VP}_e$. We re-express the best-fitting Gauss-Hermite series as a Gauss-Hermite series with zero mean (van der Marel et al. 1994b). This yields: $\sigma_e$, the dispersion of the best-fitting Gaussian to $\mathrm{VP}_e$, and $z_4$, the fourth Gauss-Hermite moment of $\mathrm{VP}_e$. The quantity $\sigma_e$ is an observational estimate of $\sigma_{\mathrm{RMS}}$, the RMS projected line-of-sight velocity. If $\mathrm{VP}_e$ is Gaussian, then $\sigma_e = \sigma_{\mathrm{RMS}}$. Figure 2 displays the $\sigma_e$ and $z_4$ profiles. If the VP is symmetric and there is no mean streaming, then $\sigma_e = \sigma$ and $z_4 = h_4$. This is the case for NGC 2434 and 2663 to within the observational errors. By contrast, in the fast rotators NGC 3706 and 5018 the $\sigma_e$ profile declines more gently in the outer parts than the $\sigma$ profile, and the $z_4$ profile differs from the $h_4$ profile.

We consider models with $\rho_L$ stratified on spheroids with axis ratio $q$, of the form:

$$\rho_L(R, z) = \rho_0 \Big(\frac{m}{b}\Big)^\alpha \Big(1 + \frac{m^2}{b^2}\Big)^\beta, \qquad m^2 = R^2 + z^2/q^2. \qquad (1)$$

The associated DFs and VPs were discussed by QZMH. The isophotes of the projected mass density have ellipticity $\epsilon = 1 - q'$, where $q'^2 \equiv \cos^2 i + q^2 \sin^2 i$, and $i$ is the inclination. For each galaxy, $\rho_0$, $b$, $\alpha$, and $\beta$ were chosen to fit the observed $R$-band surface brightness profiles, after convolution of the model with the seeing. Since we are interested in the outer regions, $\epsilon$ was fixed to the average observed value outside $0.5R_e$. It is listed in Table 1, together with the effective radii and the best-fitting parameters $b$, $\alpha$, and $\beta$. These are independent of $i$.

Since the calculation of $f(E, L_z)$ from $\rho_L$ requires substantial numerical effort, we adopted two simplifications. First, we use the Jeans equations to calculate the second velocity moments $\langle v_R^2 \rangle$ and $\langle v_\phi^2 \rangle$. Since $\langle v_R^2 \rangle = \langle v_z^2 \rangle$ and $\langle v_R v_z \rangle = 0$ in the two–integral case, $\sigma_{\mathrm{RMS}}$ follows by line–of–sight integration at projected radius $R'$. We compare the resulting $\sigma_{\mathrm{RMS}}$ to the observed $\sigma_e$. The errors thus introduced are small, because the observed $z_4$ values are relatively close to zero. Second, we only calculate $z_4$ in the limit of very large radii ($R/b \gg 1$), where its calculation is much easier (QZMH). The observed $z_4$ values extend to $R'/b \approx 5$ in NGC 2434 and 2663, and to $R'/b \approx 2$ in NGC 3706 and 5018, where $R'$ is the projected radius on the sky. We verified that these simplifications do not influence any of our main conclusions, by calculating the full DF for a number of specific cases. In particular, the $z_4$ values for the range $R'/b \approx 2 - 5$ are very close to the large radii result.

Figure 2 shows the predicted $\sigma_{\mathrm{RMS}}$ profiles for the self-consistent case (i.e., no dark halo), for edge-on models ($i = 90°$). The normalization is determined by the average mass-to-light ratio of the stellar population $M/L$, which can be chosen freely. Independent of the assumed $M/L$, we conclude the following. For NGC 2434 the predicted profile falls more steeply in the outer parts than the observed profile, while for NGC 2663 it falls less steeply. For NGC 3706 and 5018 the models predict too little motion in the outer parts relative to the inner parts. These results agree with those of CDa,b, who discussed also the effects of varying $i$ and of adding a dark halo.

The dashed curves in Figure 3 show the predicted $z_4$ values in the large radii limit as function of $i$. The dependence on $i$ is weak, except at the smallest allowed inclinations. As expected, the predicted $z_4$ is always strongly negative, because any flattened $f(E, L_z)$ model is supported mainly by azimuthal motion (e.g., Dehnen & Gerhard 1994). The hatched bands represent the error bars on $z_4$ at the outermost measured point, and are a conservative estimate of the observational uncertainty in the mean $z_4$ values measured at $R' > R_e$. Except for NGC 2663, the predicted $z_4$ values lie well below the observed values. This is not caused by the simple form adopted for $\rho_L$, and remains so when $\alpha$, $\beta$, and $b$ are varied within the range allowed by the errors of the best fit (Table 1). It also cannot be due to systematic errors in the observations, as these are $|\Delta z_4| \lesssim 0.03$ (of the same order as for $h_4$).

Models with $f(E, L_z)$ and a dark halo can fit the observed $\sigma_e$ profiles reasonably well (CDa,b). We study the $z_4$ values predicted by $f(E, L_z)$ models with a dark halo, by (non–selfconsistently) embedding the luminous mass-density $\rho_L$ in a power–law potential $\Psi_{dk} \propto (R^2 + z^2/q_{dk}^2)^{\gamma/2}$ at large radii, which corresponds to a total density profile in the equatorial plane $\rho \propto R^{\gamma-2}$ (Evans 1994). We use again the asymptotic results of QZMH for $R/b \gg 1$. In this limit $\sigma_e \propto R^{\gamma/2}$. For each galaxy, $\gamma$ was fixed by fitting to the observed $\sigma_e$ profile at large radii. The solid curves in Figure 3 show the predicted $z_4$ for different values of the axis ratio $q_{dk}$ ($= 0.8, 0.9, 1.0$) of the potential $\Psi_{dk}$. These correspond to axis ratios $q_{mdk}$ of the halo mass density ranging between 0.4 and 1.0. The predicted $z_4$ increases when $q_{dk}$ decreases, since fewer high angular momentum orbits are needed to reproduce the same flattened mass density in a flatter potential. For NGC 2434 the $z_4$ values overlap with the observed values for $q_{mdk} \approx 0.4 - 0.8$. However, for NGC 3706 and 5018 very flattened halos ($q_{mdk} \approx 0.1 - 0.4$) are required to bring the predicted $z_4$ into agreement with the observed value.

## 4. Discussion and Conclusions

The VPs of elliptical galaxies can now be measured reliably out to $\sim 2R_e$. We have presented the first such measurements for four galaxies, and have compared them to the predictions of axisymmetric $f(E, L_z)$ dynamical models. Radially anisotropic models generally predict VPs that are more centrally peaked than a Gaussian, while tangentially anisotropic models predict VPs that are more flat-topped (e.g., Gerhard 1993). The observed $z_4$ values imply a velocity distribution that is less tangentially anisotropic than in a self-consistent $f(E, L_z)$ model. However, such a velocity distribution predicts a $\sigma_e$ profile that falls off steeper than in the $f(E, L_z)$ model (e.g., Dehnen & Gerhard 1993). This is consistent with the data for NGC 2663, but it increases the discrepancy with the gently declining $\sigma_e$ profiles of NGC 2434, 3706, and 5018. Thus we conclude that for NGC 2434, 3706 and 5018, no self-consistent axisymmetric model can fit simultaneously the $\sigma_e$ and the VP data. These galaxies must have a dark halo.

We have shown that $f(E, L_z)$ models with flattened massive dark halos predict $z_4$–values in harmony with the measurements for NGC 2434, 3706 and 5018. The $f(E, L_z)$ dark halos for the latter two galaxies are very flat. Although we cannot exclude this possibility, we suspect that NGC 3706 and 5018 are surrounded by less–flattened halos. The resulting discrepancy in $z_4$ can then be explained either by small systematic errors in the $z_4$ measurements, or by an orbital structure which is less tangentially anisotropic than our $f(E, L_z)$ models. So, while for NGC 2434 possibly $f = f(E, L_z)$, for NGC 3706 and 5018 it seems more plausible that $f \neq f(E, L_z)$. To give these qualitative arguments a more quantitative basis, and to determine the size and mass of the dark halos, it will be necessary to construct flattened three-integral models for the galaxies in our sample.

We thank Martin Schwarzschild and Wyn Evans for comments on an earlier version of this paper. During part of this work, EEQ was supported by NSF grant DMS–9407559, and RPvdM was supported by NASA grant HF-1065.01-94A, awarded by the STScI which is operated by AURA, Inc., for NASA under contract NAS5-26555.

Table 1: Basic Properties and Model Parameters[a]

| Galaxy | $R_e('')$ | $\epsilon$ | $\alpha$ | $\beta$ | $b('')$ |
|--------|-----------|------------|----------|---------|---------|
| NGC 2434 | 24 | 0.08 | $-1.9$ | $-0.52$ | 13 |
| NGC 2663 | 50 | 0.30 | $-1.6$ | $-0.55$ | 10 |
| NGC 3706 | 27 | 0.35 | $-2.3$ | $-0.53$ | 31 |
| NGC 5018 | 22 | 0.30 | $-2.1$ | $-0.90$ | 31 |

[a]Listed are: the effective radius $R_e$ (Lauberts & Valentijn 1989), the average ellipticity $\epsilon$ outside $R_e/2$ (CDa,b), and the model parameters $\alpha$, $\beta$, and $b$ of equation (1). Mean errors on the best fit parameters are $\Delta\alpha = \pm0.1$, $\Delta\beta = \pm0.05$, and $\Delta b = \pm5''$.

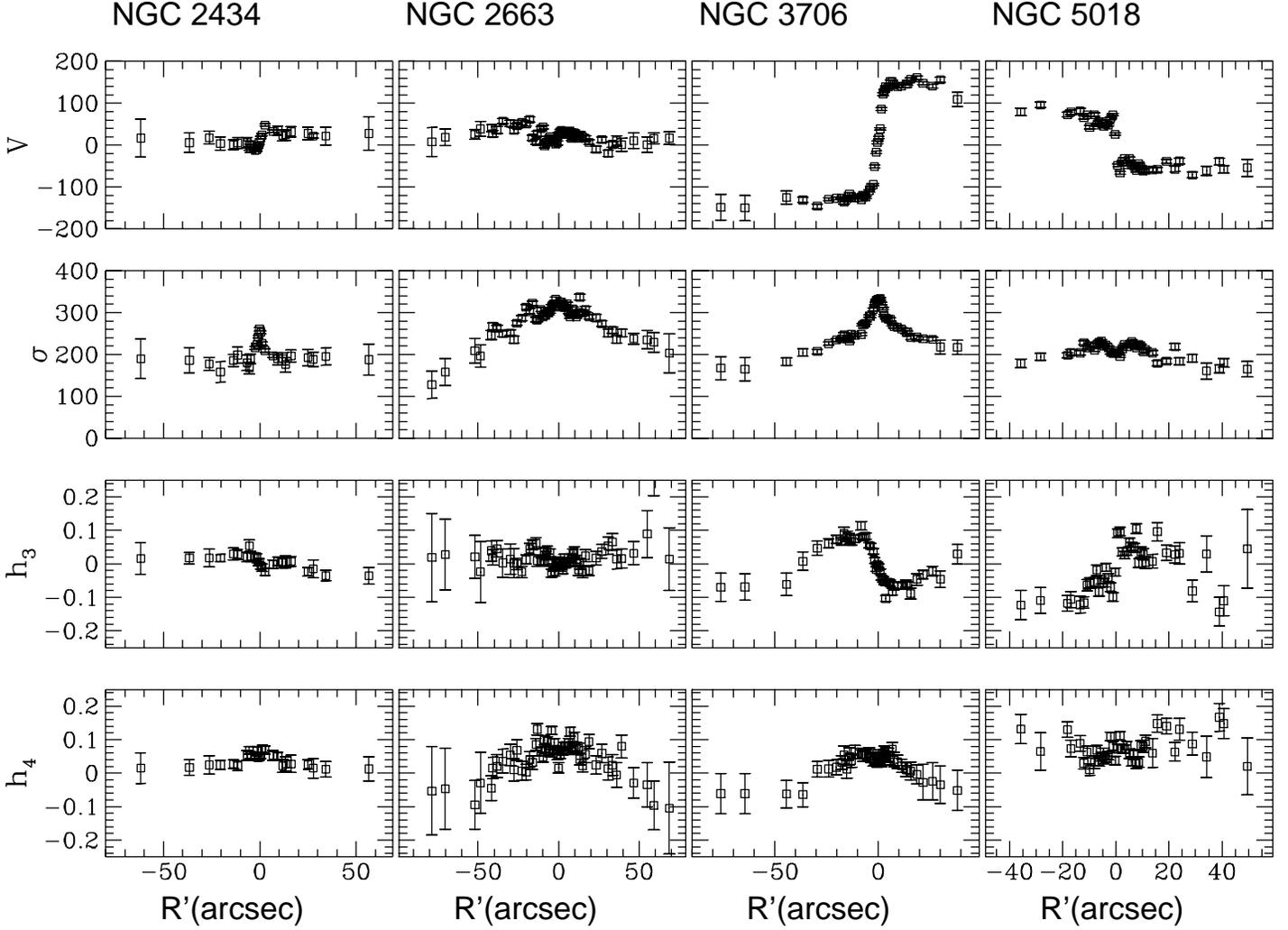

Fig. 1.— Velocity profile (VP) parameters for the four galaxies in our sample as functions of projected radius $R'$ on the plane of the sky. From top to bottom: the mean $V$ and dispersion $\sigma$ of the best-fitting Gaussian to the VP, and the Gauss–Hermite coefficients $h_3$ and $h_4$.

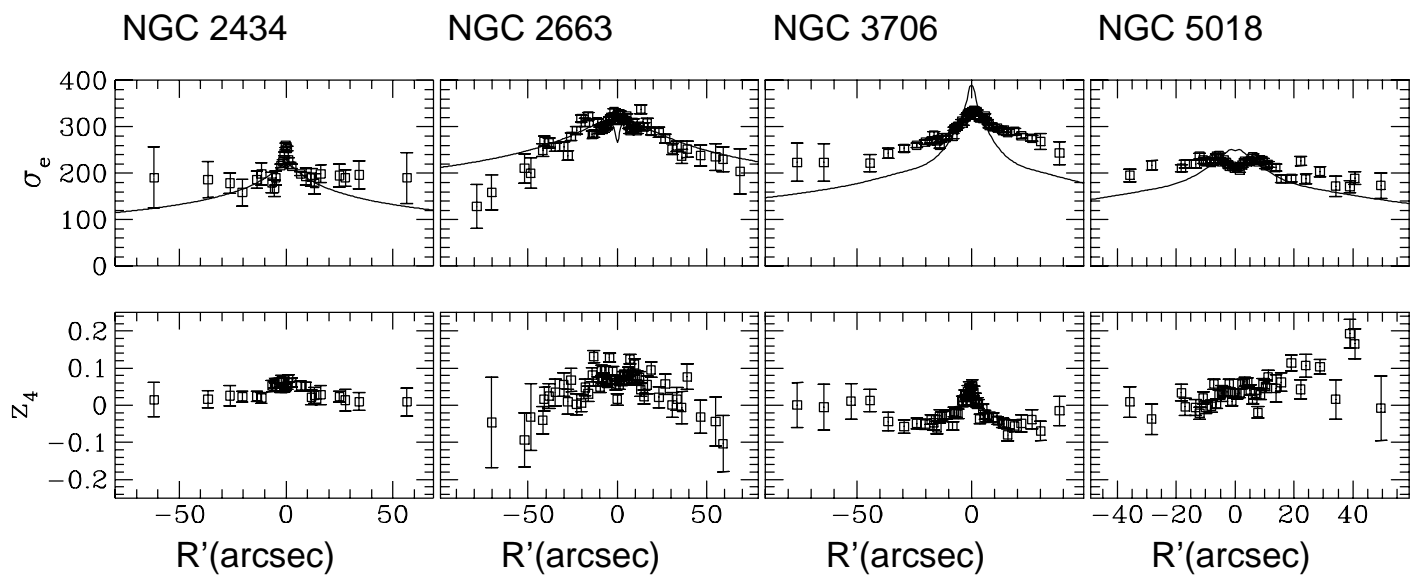

Fig. 2.— The dispersion $\sigma_e$ of the best-fitting Gaussian to the even part of the velocity profile, $VP_e$, and the Gauss-Hermite moment $z_4$ of $VP_e$, as functions of projected radius $R'$ for the four galaxies in our sample. The curves in the panels for $\sigma_e$ are the (seeing convolved) predictions for $\sigma_{RMS}$, the RMS projected line-of-sight velocity, for edge-on models with distribution function $f(E, L_z)$ and no dark halo, as described in the text.

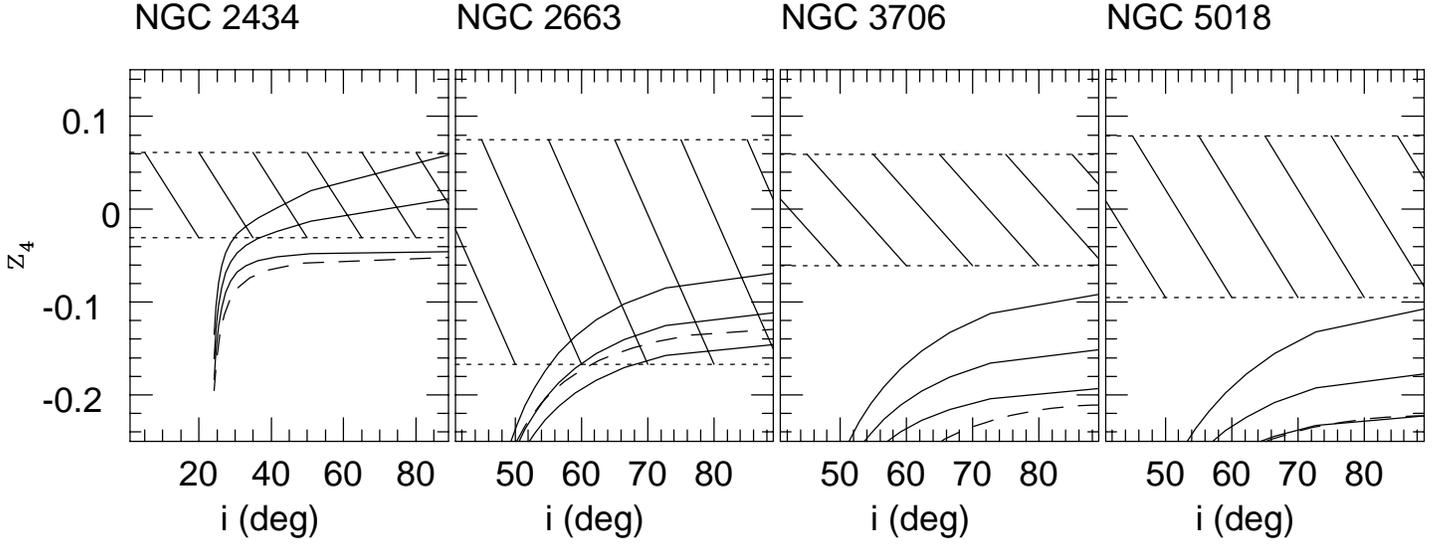

Fig. 3.— Behavior of $z_4$ as a function of inclination angle $i$ for our $(\alpha, \beta)$ models for the four galaxies in our sample. A value $z_4 < 0$ indicates that $VP_e$, the even part of the velocity profile, is more flat-topped than a Gaussian; a value $z_4 > 0$ that it is more centrally peaked. Dashed curves: $f(E, L_z)$ models without dark halo; Solid curves: $f(E, L_z)$ models with a power-law dark halo with, from top to bottom, axis ratio of the isopotential spheroids $q_{dk} = 0.8$, $0.9$, and $1.0$. The hatched regions indicate where the $z_4$ data points with $R' > R_e$ lie.